\documentclass[runningheads]{llncs}
\usepackage{graphicx}
\usepackage{caption}
\usepackage{subcaption}
\usepackage{amsmath}
\usepackage{amsfonts}

\begin{document}
\title{Generative Residual Attention Network for Disease Detection}
%
%

\author{Euyoung Kim\inst{1} \and
Soochahn Lee\inst{2} \and
Kyoung Mu Lee\inst{1}}

\authorrunning{Kim et al.}
%
\institute{Department of ECE, Automation and Systems Research Institute, Seoul National University, Seoul, Republic of Korea 
\email{\{shreka116,kyoungmu\}@snu.ac.kr}
\and
School of Electrical Engineering, Kookmin University, Seoul, Republic of Korea\\
\email{sclee@kookmin.ac.kr}}

\maketitle              
\begin{abstract}
Accurate identification and localization of abnormalities from radiology images serve as a critical role in computer-aided diagnosis (CAD) systems. Building a highly generalizable system usually requires a large amount of data with high-quality annotations, including disease-specific global and localization information. However, in medical images, only a limited number of high-quality images and annotations are available due to annotation expenses. In this paper, we explore this problem by presenting a novel approach for disease generation in X-rays using a conditional generative adversarial learning. Specifically, given a chest X-ray image from a source domain, we generate a corresponding radiology image in a target domain while preserving the identity of the patient. We then use the generated X-ray image in the target domain to augment our training to improve the detection performance. We also present a unified framework that simultaneously performs disease generation and localization.We evaluate the proposed approach on the X-ray image dataset provided by the Radiological Society of North America (RSNA), surpassing the state-of-the-art baseline detection algorithms.

\end{abstract}
\section{Introduction}

It is almost certain that radiologists know what a normal chest X-ray (CXR), that is, an image without any disease, would look like for a particular patient. While it is not fully understood how radiologists detect diseases from CXR, it is likely that they use their image of a normal one as a reference. We present a novel method for disease detection based on this premise, that the difference, or residual, between corresponding normal and abnormal images is a key tool for detecting disease regions.

The problem is that the corresponding image pair required to compute this residual often does not exist, and only either the normal or abnormal is available. We thus develop a method to construct corresponding normal-abnormal CXR image pseudo-pairs, aided by recent advances on generative methods~\cite{antoniou2017data,bowles2018gan,shrivastava2017learning}. This generation framework has an additional benefit, in that it can be easily applied to data augmentation. 

\begin{figure}[t]
\centering
\includegraphics[width=\linewidth]{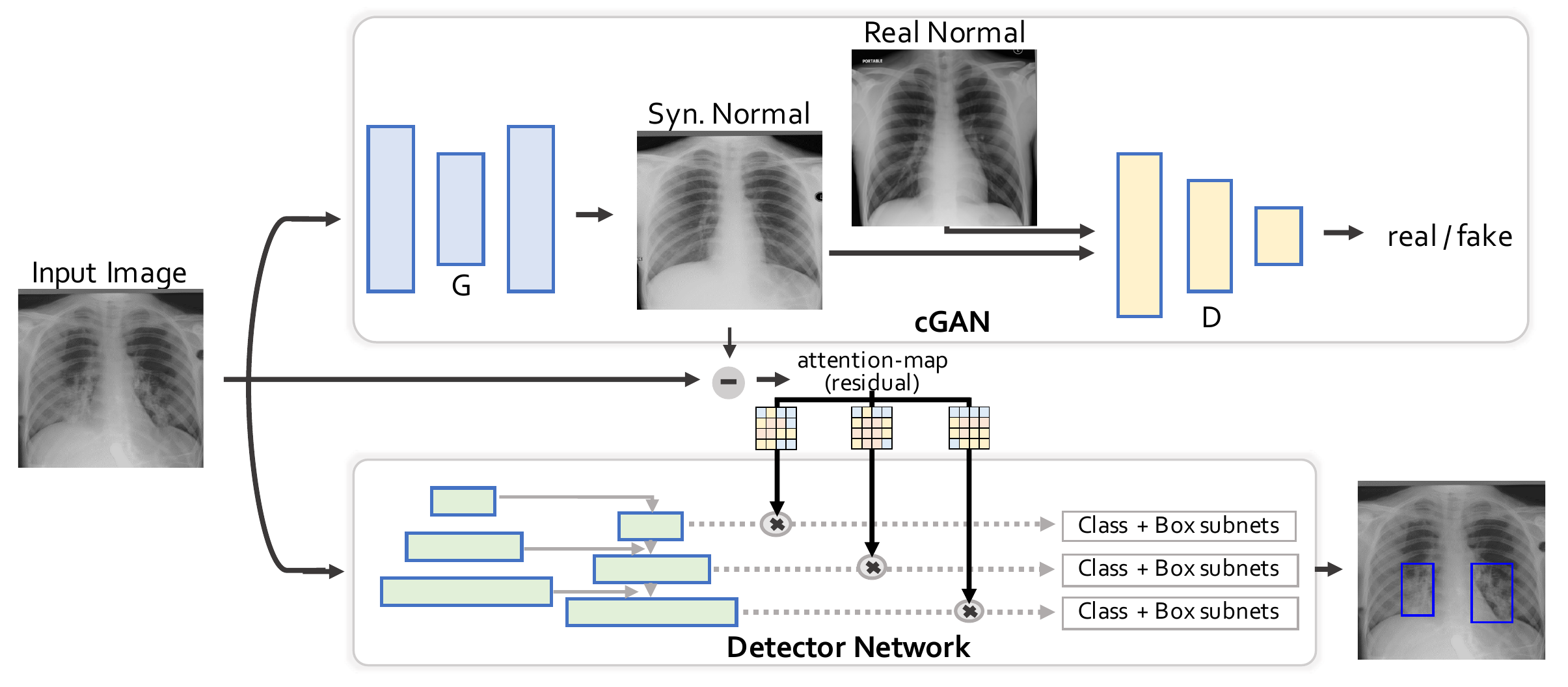}
\caption{Structure of the proposed generative residual attention network (GRANet).}
\label{fig:GRANet}
\end{figure}

The key contributions of our method are three-fold. First, we propose the generative residual attention network (GRANet) model, which applies the residual between the generated corresponding {\emph normal} image and the input image as the attention map within conventional detection models as depicted in Fig.~\ref{fig:GRANet}. Second, we propose a new framework for generating corresponding pseudo image pairs required for supervised learning of the GRANet. Third, we propose a method for data augmentation using the framework of generating pseudo image pairs.

In general, our method is related to previous methods using generative adversarial learning for CXR pathology classification~\cite{salehinejad2018generalization,madani2018chest,tang2019abnormal}.
Perhaps it is more closely related to the works of Liu et al.~\cite{Liu_2019_ICCV} and Tang et al.~\cite{tang2019xlsor,tang2021disentangled}, which present methods for generating CXR images mostly in the context of data augmentation. But our formulation of the residual attention results in key differences in the generation process.

The CXR has been one of the most widely adopted diagnostic imaging tests for various thoracic diseases including pneumonia, cardiomegaly, and cancer. In this paper, we evaluate the efficacy of the proposed method on the detection of pneumonia. Experimental results demonstrate that both the residual attention and the data augmentation contribute to improvements in average precision. As the proposed method should not be limited to a particular disease, we believe that it can also be generalized to various diseases to improve computer-aided diagnosis systems for CXRs.

\section{Method}

\subsection{Generative Residual Attention Network}

Essentially, the GRANet is a conventional detector with an additional generative branch, in which its output is used as the attention map for the detector features. As depicted in Fig.~\ref{fig:GRANet}, the input image is fed into a conditional GAN (cGAN)~\cite{isola2017image,zhu2017unpaired} that generates a corresponding normal CRX image $G(x)$. The difference between the two, or the residual image $I_{r} = |x - G(x)|$, is downsampled to the feature dimensions and used as the spatial attention map. If the input image contains pathological regions, the residual image should only be non-zero at those pathological regions, and thus enforce the detector network to attend to those regions. If the input image is a normal image, residual should be zero, i.e., cGAN should act as an identity. To train the GRANet, we first train the cGAN on pseudo CRX pairs, and then alternately train the detector network and the cGAN.

\subsection{Normal Image Generation for Supervised Learning of GRANet} \label{ABN2NOR_GEN}

Our framework for constructing of pseudo-normal CRX image for a given abnormal image comprises the following components: 1) image alignment, 2) replacing pathological bounding box region with nearest normal neighbor, 3) Poisson image blending, and 4) radio-realistic pseudo-normal generation. Our idea is to replace the pathological bounding box region in the given abnormal image with that of a similar normal image. This blended image will only have non-zero residual values within the ground truth bounding box. These pseudo-pair images are then used to train the cGAN, which is the initial cGAN used in the GRANet. Next, we describe the components in detail.

\begin{figure}[!ht]
\begin{center}
   \includegraphics[width=\linewidth]{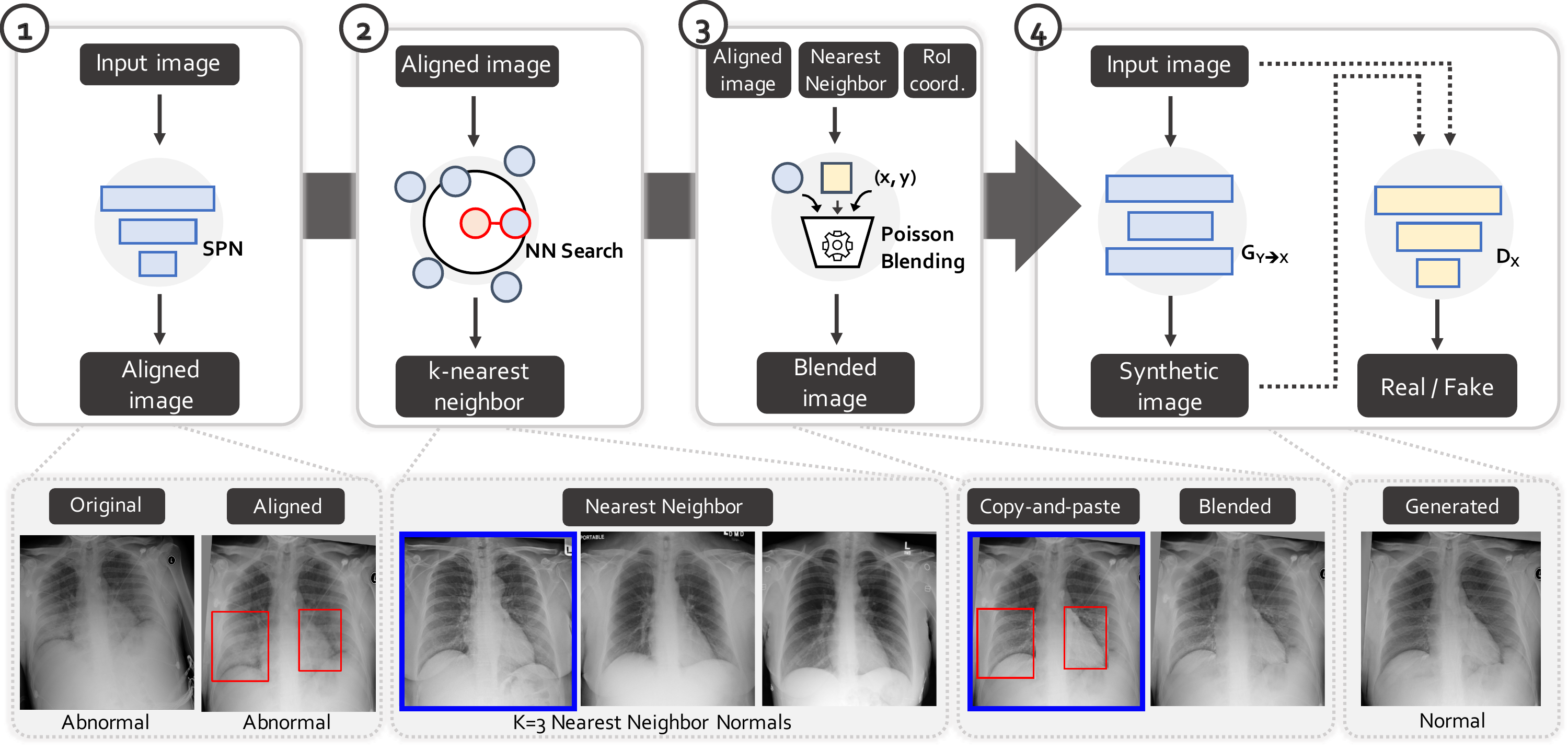}
\end{center}
   \caption{Framework for generating pseudo-normal image corresponding to the input abnormal image for supervised learning of the GRANet.}
\label{fig:translation}
\end{figure}

\noindent
{\bf Alignment Module:}
Just like face images, CXRs also share similar structures and can thus be aligned to a single reference image. We formulate the problem of alignment as transformation learning by minimizing the structure divergence between the transformed image and the reference image. We manually select a standard high-quality image and set it as the reference image, and train the spatial transformer network (STN)~\cite{jaderberg2015spatial}. 

Given input and reference images $I$ and $R$, affine transform matrix $\theta$, and the STN $T_\theta$, the loss function is defined as the sum of the feature reconstruction in perceptual loss~\cite{johnson2016perceptual} $f$ and the consistency loss measured as the L2 distance between transformed and input images: 
\begin{equation} \label{stn_eqn}
L_\theta = f(T_\theta(I), R) + ||T_\theta(I)- R||_{2}.
\end{equation}
We use ResNet18~\cite{he2016deep} as the backbone of $T_\theta$.

\noindent
{\bf Replacing Disease Region with Nearest Normal Neighbor:}
For each abnormal CXR we find a structurally similar normal CXR to construct pairs. We find that using simple nearest neighbor search algorithm is good enough for this task over other algorithms such as perceptual hashing. Specifically, we first resize all images to 32x32, then for each resized abnormal image, we choose the nearest normal image based on euclidean distance. Then, we copy and paste pixel values from this normal image into the given input image for the bounding box region.

\noindent
{\bf Poisson Blending:}
To reduce the boundary artifacts of the copied bounding regions, we adopt Poisson blending~\cite{perez2003poisson}, which smooths the abrupt intensity changes at the boundary to make a more visually realistic looking composition. For clarity, we denote this blended image as $I_b$. Although $I_b$ can be used as the pseudo-normal image, it may be inconsistent with real CXR images, considering the different origin of the local regions blended into the given input image.

\noindent
{\bf Radio-realistic Pseudo-normal Generation:} To generate the final radio-realistic pseudo-normal CXR, we apply a conditional generative learning with $I_{b}$ as input images. That is, the blended $I_{b}(x)$ images for all abnormal image $x$ are used as inputs to the generator $G$ of a cGAN, while all real normal image $y$ are used as inputs for the discriminator $D$. 
The objective function is represented as follows:
\begin{equation} \label{adv_eqn}
\begin{split}
    \mathcal{L}_{realistic} & = \mathbb{E}_{y\sim p_{data}(y)}[log D(y)] + \mathbb{E}_{x\sim p_{data}(x)}[log (1-D(G(I_{b}(x)))]\\ & 
                          + ||I_{b}(x)-G(I_{b}(x))||_1,
\end{split}
\end{equation}
where the first two terms represent the adversarial loss to ensure realistic generation, and the third term represents the reconstruction loss to minimize changes with $I_{b}$.
After training this cGAN, the resulting $G_{realistic}(I_b(x))$ is defined as the pseudo-normal image for input abnormal image $x$.

\noindent
{\bf Training the Abnormal-to-Normal Generator of the GRANet:} After generating the corresponding pseudo-normal images for the abnormal images, we then train the cGAN of the GRANet, to generate normal images from abnormal image inputs. We actually use the same cGAN structure with same normal real images, but only with different input images, the original abnormal image $x$ instead of the blended $I_{b}(x)$, to the generator. 
Thus, the objective function is represented as follows:
\begin{equation}
\begin{split}
    \mathcal{L}_{Abn2Nor} & = \mathbb{E}_{y\sim p_{data}(y)}[log D(y)] + \mathbb{E}_{x\sim p_{data}(x)}[log (1-D(G(x))] \\
                          & + ||G_{realistic}(I_b(x))-G(x)||_1.
\end{split}                                                  
\end{equation}

\subsection{Abnormal Image Generation for Data Augmentation} \label{NOR2ABN_GEN}

We apply the same framework used to generate pseudo-normal images but instead to generate pseudo-abnormal images. We only need to assign normal images as the input, search for nearest neighbor abnormal images, and replace normal regions with pathological regions. We can thus generate many pseudo-abnormal images for data augmentation which are fed into the GRANet as inputs along with the real images.

\section{Implementation}
\subsection{Network Architecture}
\noindent
{\bf Generative Model:}
For our generative model, we adopt the architecture from CycleGAN~\cite{zhu2017unpaired}. The generator consists of three convolutions followed by 9 residual blocks~\cite{he2016deep}, two fractionally-strided convolutions with stride 2, and a final convolution that maps features to the CXR image. For the discriminator, we adopt the architecture of 70x70 PatchGANs~\cite{demir2018patch,isola2017image,ledig2017photo}, since such a patch-level discriminator has fewer parameters than a full-image discriminator, while being applicable to arbitrarily sized images in a fully convolutional fashion.

\noindent
{\bf Detection Model:}
We perform experiments with two different baseline detectors, each from single- and two-stage detection streams, namely, the Mask-RCNN~\cite{he2017mask} and RetinaNet~\cite{lin2017focal}. As in \cite{gabruseva2020deep}, we make the following modifications to RetinaNet: 1) use SE-ResNeXt101~\cite{hu2018squeeze} as backbone, 2) add loss for negative samples, 3) add pyramid level for smaller anchors, 4) add global label classification module on top of box regression and classification, and 5) addition of dropout.

\subsection{Training details}

For the training of cGAN in radio-realistic pseudo-normal generation, we train the model from scratch, with Adam optimizer, initial learning rate 0.0002, and batch size 4, on a GTX 1080Ti. 

For the training of cGAN in the GRANet, we follow the configuration in \cite{gabruseva2020deep}. Adam optimizer\cite{kingma2014adam}, initial learning rate 0.00001, batch size 6, and standard data augmentation including affine transformation, horizontal flip, Gaussian noise and blur, etc. Weights are initialized by pre-trained SE-ResNeXt101 on ImageNet~\cite{deng2009imagenet}. 

In generating pseudo-abnormal images for data augmentation, for each image that does not contain pneumonia, we search for $k=1$ nearest neighbor abnormal image and replace regions with pathological regions with pneumonia resulting in about $20,000$ additional pseudo-abnormal images to the original dataset. Note that as the number of $k$ should not be limited to 1, theoretically we can further increase the number of pseudo-abnormal images by searching for $k>1$ nearest neighbors, if needed.

\section{Experiments}
\subsection{Dataset and Evaluation Metrics}
{\bf Dataset:}
We evaluate on the public pneumonia chest X-rays dataset~\cite{wang2017chestx,shih2019augmenting} provided by the RSNA~\cite{rsna} which includes disease labels along with the region-level bounding boxes annotations. The dataset comprises frontal-view CXRs from 26,684 unique patients, where each image is labelled with one of the following classes from the associated radiological reports: 1) normal (8851), 2) lung opacity (6012), and 3) no lung opacity but not normal. 

We conduct experiments in two different settings. In the first, we use the same splits as in \cite{gabruseva2020deep} for comparative evaluation. 4-fold cross-validation is performed on the data, which is split at the patient-level. In the second, a holdout set of 600 images were used for testing and the rest were used for training, as in \cite{yao2020improved}. For all experiments, we downsample the original CXRs of $1024 \times 1024$ resolution to $512 \times 512$ due to computational constraints.

\noindent
{\bf Evaluation Metrics:}   
We evaluate the model using the average precision (AP) following the metrics used in the Kaggle pneumonia detection challenge~\cite{rsna}. AP is measured at different intersection-over-union (IoU) thresholds. The threshold values $t$ range from $0.4$ to $0.75$ with a step size of $0.05$ for the first dataset settings and from $0.4$ to $0.6$ with step size $0.1$ for the second dataset settings. The average precision of a single image is calculated as the mean of the precision values at each IoU threshold:

\begin{equation}
   \textbf{mAP} = \frac{1}{|thresholds|}\sum_t\frac{TP(t)}{TP(t)+FP(t)+FN(t)}.
   \label{eqn:metric}
\end{equation}
\noindent
A true positive (TP) is counted when a single predicted object matches a ground truth object with an IoU above the threshold. A false positive (FP) indicates a predicted object had no associated ground truth object. A false negative (FN) indicates a ground truth object had no associated predicted object. Note that in the second experiment setting, the average precision is measured with the equation~\ref{eqn:metric} without counting false negatives (FN).

\subsection{Comparative Evaluation}
\begin{figure}[!tbh]
\begin{center}
   \includegraphics[width=\linewidth]{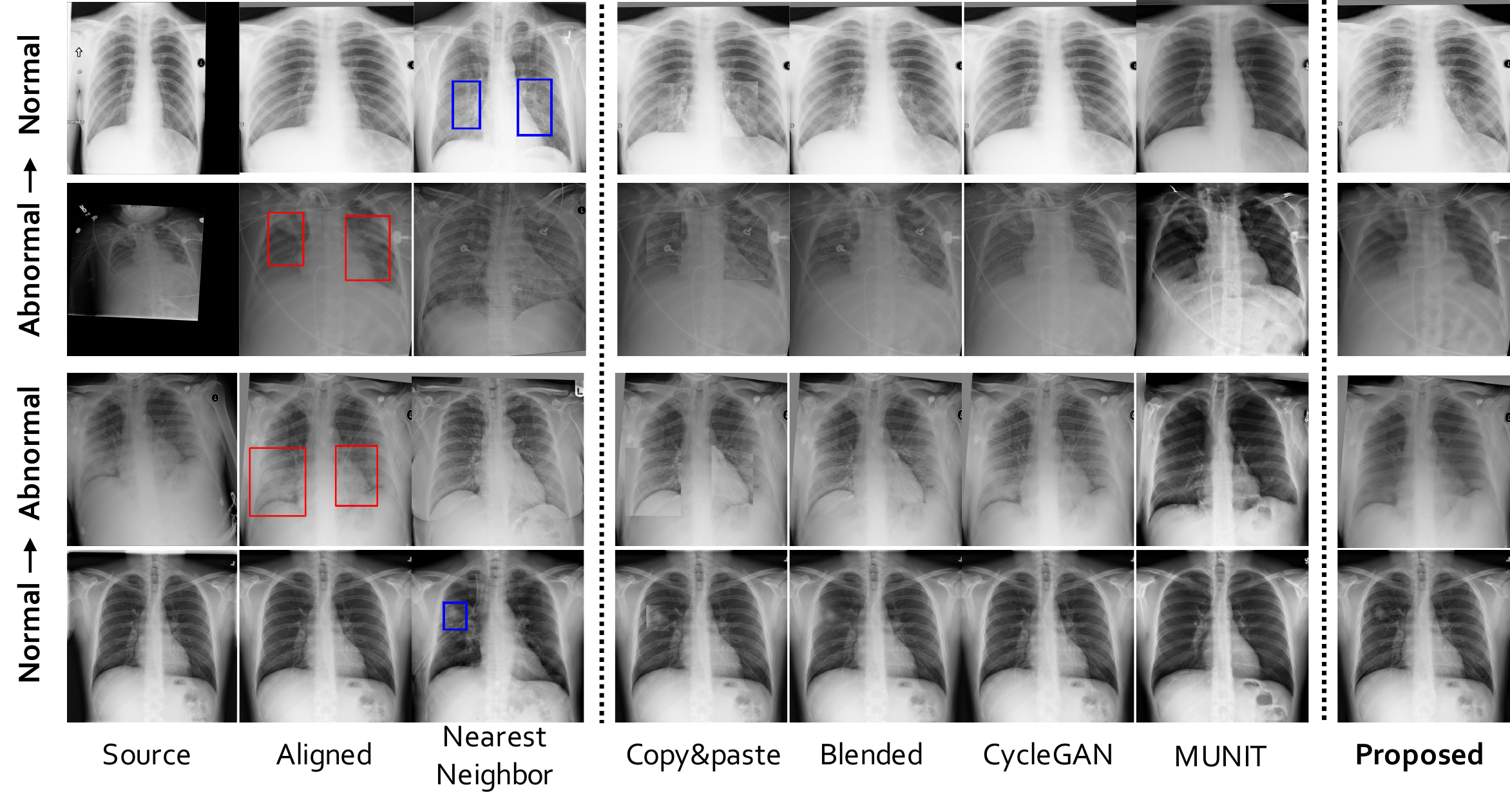}
\end{center}
   \caption{Four examples of the generated normal and abnormal CXRs.}
   \label{fig:qualitative}
\end{figure}

\begin{table}[!tbh]
\begin{center}
\setlength{\tabcolsep}{16pt}
\begin{tabular}{l c c c c c}
\hline
Method & \multicolumn{1}{l}{AP}  & \multicolumn{1}{l}{AP$_{0.4}$}   & \multicolumn{1}{l}{AP$_{0.5}$}   & \multicolumn{1}{l}{AP$_{0.6}$}  \\
\hline\hline
RetinaNet$^{\ast}$~\cite{lin2017focal}   & 0.191 & 0.414 & 0.142 & 0.127 \\
+ DGM~\cite{tang2021disentangled} & 0.239 & 0.446 & 0.242 & 0.166 \\
\hline
RetinaNet$^{\dagger}$~\cite{gabruseva2020deep}& 0.239          & 0.355 & 0.304 & 0.233\\
\textbf{+ align}                        & \textbf{0.240} & \textbf{0.359} & \textbf{0.309} & \textbf{0.231}     \\
\textbf{+ align, attn.}                 & \textbf{0.242} & \textbf{0.367} & \textbf{0.312} & \textbf{0.234}     \\
\textbf{+ align, syn.}                  & \textbf{0.256} & \textbf{0.366} & \textbf{0.320} & \textbf{0.244}     \\
\textbf{+ align, attn. , syn.}          & \textbf{0.260} & \textbf{0.373} & \textbf{0.326} & \textbf{0.254}     \\
\hline

\end{tabular}
\end{center}
\caption{Comparison of pneumonia detection results with different training augmentation configurations in terms of average precision (AP) used in the challenge. AP: mean average precision@IoU=[0.4:0.05:0.75], AP$_{0.4}$ : AP@IoU=0.4, and so forth. RetinaNet$^{\ast}$ uses ResNet50, while RetinaNet$^{\dagger}$ uses SE\_ResNeXt101 as a backbone.}
\label{tbl:single-stage}
\end{table}
{\bf CXR Generation:}
We first compare the generation quality with the seminal works on unsupervised image-to-image translations~\cite{zhu2017unpaired,huang2018multimodal} since both works are capable of synthesizing images in either from normal or abnormal images. We demonstrate the efficacy of the proposed approach in both tasks of generating abnormal and normal CXRs in Fig.~\ref{fig:qualitative}. Our cGAN is capable of generating normal images from abnormal images, and vice-versa, while other two approaches were unable to achieve compelling results. Moreover, in abnormal generation, our cGAN is able to generate pathological regions in a specific location.

\noindent
{\bf Pneumonia Detection:}
We perform quantitative comparisons with three recent state-of-the-art methods~\cite{gabruseva2020deep,tang2021disentangled,yao2020improved}. Comparative evaluation of the modified RetinaNet$^\dagger$~\cite{gabruseva2020deep}, the disentangled generative model (DGM)~\cite{tang2021disentangled}, and the proposed method based on RetinaNet$^\dagger$ is presented in Table~\ref{tbl:single-stage}, with the dataset setting following \cite{gabruseva2020deep}. Comparative evaluation of Faster RCNN~\cite{ren2015faster}, SSD~\cite{liu2016ssd}, YOLO~\cite{redmon2018yolov3}, PYolo~\cite{yao2020improved}, and the proposed method based on Mask-RCNN~\cite{he2017mask} is presented in Table~\ref{tbl:two-stage}, with the dataset setting following \cite{yao2020improved}. We observe consistent improvements on the baseline by augmenting the training with pseudo-abnormal generation and residual attention.

\noindent
{\bf Ablation Study}
We perform an ablation study on different settings in training strategy to train the cGAN and the detection model, simultaneously. As shown in Table~\ref{tbl:ablation}, we observe that iterative training scheme outperforms joint training of both the generative and detection models concurrently. Proposed iterative scheme alternately updates one model for $n$ epochs while freezing the parameters of other model. For example, the 2:2 iterative scheme first updates the detector network for two epochs while freezing the generative network, then for the next two epochs the generative network is subjected to update while the parameters of the detector network are unchanged. We observe that the models trained with evenly distributed update iterations for each model consistently outperform over models trained with one-sided iterative schemes.

\begin{table}[t]
\begin{center}
\setlength{\tabcolsep}{10pt}
\begin{tabular}{l c c c c}
\hline
Method                   & \multicolumn{1}{l}{AP}  & \multicolumn{1}{l}{AP$_{0.4}$}   & \multicolumn{1}{l}{AP$_{0.5}$}   & \multicolumn{1}{l}{AP$_{0.6}$} \\
\hline\hline
Faster RCNN~\cite{ren2015faster}           & 0.430 & 0.586 & 0.438 & 0.266 \\
SSD~\cite{liu2016ssd}                      & 0.404 & 0.559 & 0.406 & 0.247 \\
Yolo\_v3~\cite{redmon2018yolov3}           & 0.434 & 0.579 & 0.430 & 0.292 \\
PYolo\_v3~\cite{yao2020improved}           & 0.468 & 0.642 & 0.447 & 0.316 \\
\hline
\textbf{Mask RCNN~\cite{he2017mask} + align}        & \textbf{0.476} & \textbf{0.645} & \textbf{0.504} & \textbf{0.279} \\
\textbf{Mask RCNN~\cite{he2017mask} + align + attn.}& \textbf{0.520} & \textbf{0.695} & \textbf{0.554} & \textbf{0.311} \\
\hline

\end{tabular}
\end{center}
\caption{Comparison of performance with various detection frameworks in terms of conventional average precision (AP). AP: mean average precision@IoU=[0.4:0.1:0.6], AP$_{0.4}$ : AP@IoU=0.4, and so forth.}
\label{tbl:two-stage}
\end{table}
\begin{table}[t]
\begin{center}
\setlength{\tabcolsep}{8pt}
\begin{tabular}{l | c c c c c c c c}
\hline
Strategy   & joint & 1:1   & 1:2   & 1:3   & 2:1   & 2:2   & 3:3   & 5:1 \\
\hline\hline
AP     & 0.222 & 0.240 & 0.245 & 0.233 & 0.237 & \textbf{0.246} & 0.242 & 0.233 \\
\hline

\end{tabular}
\end{center}
\caption{Ablation study on different training strategy configurations in GRANet.}
\label{tbl:ablation}
\end{table}

\section{Conclusion}
In this paper, we investigate how the generative model can be adopted to the disease detection framework to improve the detection performance. We propose a novel generative residual attention network along with radiorealisitc CXR generation framework. Experiments demonstrate that the proposed generative model can synthesize radiorealistic CXRs in both abnormal and normal domains. In addition, augmenting the training of detection framework via residual attention further improves the detection performances. Future work will explore multi-class multi-label generative model that can translate multiple disease with location specific information.

\section*{Acknowledgement}
This work was supported by the Interdisciplinary Research Initiatives Program from College of Engineering and College of Medicine, Seoul National University (grant no. 800-20170169) and the National Research Foundation of Korea (NRF) grants funded by the Korean government (MSIT) (No. NRF- 2017R1A2B2011862).

%
%
%
\bibliographystyle{splncs04}
\bibliography{egbib}

\begin{thebibliography}{10}
\providecommand{\url}[1]{\texttt{#1}}
\providecommand{\urlprefix}{URL }
\providecommand{\doi}[1]{https://doi.org/#1}

\bibitem{antoniou2017data}
Antoniou, A., Storkey, A., Edwards, H.: Data augmentation generative
  adversarial networks. arXiv preprint arXiv:1711.04340  (2017)

\bibitem{bowles2018gan}
Bowles, C., Chen, L., Guerrero, R., Bentley, P., Gunn, R., Hammers, A., Dickie,
  D.A., Hern{\'a}ndez, M.V., Wardlaw, J., Rueckert, D.: Gan augmentation:
  Augmenting training data using generative adversarial networks. arXiv
  preprint arXiv:1810.10863  (2018)

\bibitem{rsna}
Challenge, R.P.D.: Covid19 radiological society of north america (2018),
  https://www.kaggle.com/c/rsna-pneumonia-detection-challenge

\bibitem{demir2018patch}
Demir, U., Unal, G.: Patch-based image inpainting with generative adversarial
  networks. arXiv preprint arXiv:1803.07422  (2018)

\bibitem{deng2009imagenet}
Deng, J., Dong, W., Socher, R., Li, L.J., Li, K., Fei-Fei, L.: Imagenet: A
  large-scale hierarchical image database. In: 2009 IEEE conference on computer
  vision and pattern recognition. pp. 248--255. Ieee (2009)

\bibitem{gabruseva2020deep}
Gabruseva, T., Poplavskiy, D., Kalinin, A.: Deep learning for automatic
  pneumonia detection. In: Proceedings of the IEEE/CVF Conference on Computer
  Vision and Pattern Recognition Workshops. pp. 350--351 (2020)

\bibitem{he2017mask}
He, K., Gkioxari, G., Doll{\'a}r, P., Girshick, R.: Mask r-cnn. In: Proceedings
  of the IEEE international conference on computer vision. pp. 2961--2969
  (2017)

\bibitem{he2016deep}
He, K., Zhang, X., Ren, S., Sun, J.: Deep residual learning for image
  recognition. In: Proceedings of the IEEE conference on computer vision and
  pattern recognition. pp. 770--778 (2016)

\bibitem{hu2018squeeze}
Hu, J., Shen, L., Sun, G.: Squeeze-and-excitation networks. In: Proceedings of
  the IEEE conference on computer vision and pattern recognition. pp.
  7132--7141 (2018)

\bibitem{huang2018multimodal}
Huang, X., Liu, M.Y., Belongie, S., Kautz, J.: Multimodal unsupervised
  image-to-image translation. In: Proceedings of the European Conference on
  Computer Vision (ECCV). pp. 172--189 (2018)

\bibitem{isola2017image}
Isola, P., Zhu, J.Y., Zhou, T., Efros, A.A.: Image-to-image translation with
  conditional adversarial networks. In: Proceedings of the IEEE conference on
  computer vision and pattern recognition. pp. 1125--1134 (2017)

\bibitem{jaderberg2015spatial}
Jaderberg, M., Simonyan, K., Zisserman, A., et~al.: Spatial transformer
  networks. In: Advances in neural information processing systems. pp.
  2017--2025 (2015)

\bibitem{johnson2016perceptual}
Johnson, J., Alahi, A., Fei-Fei, L.: Perceptual losses for real-time style
  transfer and super-resolution. In: European conference on computer vision.
  pp. 694--711. Springer (2016)

\bibitem{kingma2014adam}
Kingma, D.P., Ba, J.: Adam: A method for stochastic optimization. arXiv
  preprint arXiv:1412.6980  (2014)

\bibitem{ledig2017photo}
Ledig, C., Theis, L., Husz{\'a}r, F., Caballero, J., Cunningham, A., Acosta,
  A., Aitken, A., Tejani, A., Totz, J., Wang, Z., et~al.: Photo-realistic
  single image super-resolution using a generative adversarial network. In:
  Proceedings of the IEEE conference on computer vision and pattern
  recognition. pp. 4681--4690 (2017)

\bibitem{lin2017focal}
Lin, T.Y., Goyal, P., Girshick, R., He, K., Doll{\'a}r, P.: Focal loss for
  dense object detection. In: Proceedings of the IEEE international conference
  on computer vision. pp. 2980--2988 (2017)

\bibitem{Liu_2019_ICCV}
Liu, L., Muelly, M., Deng, J., Pfister, T., Li, L.J.: Generative modeling for
  small-data object detection. In: Proceedings of the IEEE/CVF International
  Conference on Computer Vision (ICCV) (October 2019)

\bibitem{liu2016ssd}
Liu, W., Anguelov, D., Erhan, D., Szegedy, C., Reed, S., Fu, C.Y., Berg, A.C.:
  Ssd: Single shot multibox detector. In: European conference on computer
  vision. pp. 21--37. Springer (2016)

\bibitem{madani2018chest}
Madani, A., Moradi, M., Karargyris, A., Syeda-Mahmood, T.: Chest x-ray
  generation and data augmentation for cardiovascular abnormality
  classification. In: Medical Imaging 2018: Image Processing. vol. 10574, p.
  105741M. International Society for Optics and Photonics (2018)

\bibitem{perez2003poisson}
P{\'e}rez, P., Gangnet, M., Blake, A.: Poisson image editing. In: ACM SIGGRAPH
  2003 Papers, pp. 313--318 (2003)

\bibitem{redmon2018yolov3}
Redmon, J., Farhadi, A.: Yolov3: An incremental improvement. arXiv preprint
  arXiv:1804.02767  (2018)

\bibitem{ren2015faster}
Ren, S., He, K., Girshick, R., Sun, J.: Faster r-cnn: Towards real-time object
  detection with region proposal networks. In: Advances in neural information
  processing systems. pp. 91--99 (2015)

\bibitem{salehinejad2018generalization}
Salehinejad, H., Valaee, S., Dowdell, T., Colak, E., Barfett, J.:
  Generalization of deep neural networks for chest pathology classification in
  x-rays using generative adversarial networks. In: 2018 IEEE International
  Conference on Acoustics, Speech and Signal Processing (ICASSP). pp. 990--994.
  IEEE (2018)

\bibitem{shih2019augmenting}
Shih, G., Wu, C.C., Halabi, S.S., Kohli, M.D., Prevedello, L.M., Cook, T.S.,
  Sharma, A., Amorosa, J.K., Arteaga, V., Galperin-Aizenberg, M., et~al.:
  Augmenting the national institutes of health chest radiograph dataset with
  expert annotations of possible pneumonia. Radiology: Artificial Intelligence
  \textbf{1}(1),  e180041 (2019)

\bibitem{shrivastava2017learning}
Shrivastava, A., Pfister, T., Tuzel, O., Susskind, J., Wang, W., Webb, R.:
  Learning from simulated and unsupervised images through adversarial training.
  In: Proceedings of the IEEE conference on computer vision and pattern
  recognition. pp. 2107--2116 (2017)

\bibitem{tang2019xlsor}
Tang, Y., Tang, Y., Xiao, J., Summers, R.M.: Xlsor: A robust and accurate lung
  segmentor on chest x-rays using criss-cross attention and customized
  radiorealistic abnormalities generation. arXiv preprint arXiv:1904.09229
  (2019)

\bibitem{tang2021disentangled}
Tang, Y., Tang, Y., Zhu, Y., Xiao, J., Summers, R.M.: A disentangled generative
  model for disease decomposition in chest x-rays via normal image synthesis.
  Medical Image Analysis  \textbf{67},  101839 (2021)

\bibitem{tang2019abnormal}
Tang, Y.X., Tang, Y.B., Han, M., Xiao, J., Summers, R.M.: Abnormal chest x-ray
  identification with generative adversarial one-class classifier. In: 2019
  IEEE 16th International Symposium on Biomedical Imaging (ISBI 2019). pp.
  1358--1361. IEEE (2019)

\bibitem{wang2017chestx}
Wang, X., Peng, Y., Lu, L., Lu, Z., Bagheri, M., Summers, R.M.: Chestx-ray8:
  Hospital-scale chest x-ray database and benchmarks on weakly-supervised
  classification and localization of common thorax diseases. In: Proceedings of
  the IEEE conference on computer vision and pattern recognition. pp.
  2097--2106 (2017)

\bibitem{yao2020improved}
Yao, S., Chen, Y., Tian, X., Jiang, R., Ma, S.: An improved algorithm for
  detecting pneumonia based on yolov3. Applied Sciences  \textbf{10}(5), ~1818
  (2020)

\bibitem{zhu2017unpaired}
Zhu, J.Y., Park, T., Isola, P., Efros, A.A.: Unpaired image-to-image
  translation using cycle-consistent adversarial networks. In: Proceedings of
  the IEEE international conference on computer vision. pp. 2223--2232 (2017)

\end{thebibliography}

\end{document}